\documentclass[amsmath,amssymb,aps,prl,twocolumn,superscriptaddress]{revtex4-1}

\usepackage{graphicx}
\usepackage{dcolumn}
\usepackage{bm}
\usepackage[T1]{fontenc}
\usepackage{lmodern}
\usepackage{floatrow}%
\usepackage{cancel}
\usepackage[caption=false]{subfig}
\usepackage{braket}

\begin{document}

\title{Quantum limit to nonequilibrium heat-engine performance imposed by strong system-reservoir coupling}

\author{David Newman}
\affiliation{Department of Physics and Astronomy, The University of Manchester, Oxford Road, Manchester, M13 9PL, UK}
\affiliation{Department of Physics, Imperial College London, London, SW7 2AZ, UK}

\author{Florian Mintert}
\affiliation{Department of Physics, Imperial College London, London, SW7 2AZ, UK}

\author{Ahsan Nazir}\email{ahsan.nazir@manchester.ac.uk}
\affiliation{Department of Physics and Astronomy, The University of Manchester, Oxford Road, Manchester, M13 9PL, UK}

\date{\today}

\begin{abstract}
We show that finite 
system-reservoir coupling imposes a distinct quantum limit on the performance of a non-equilibrium 
quantum heat engine. 
Even in the absence of quantum friction along the isentropic strokes, 
finite system-reservoir coupling induces correlations  
that result in the generation of coherence  
between the energy eigenstates 
of the working system. This coherence acts to hamper the engine's power output, as well as the efficiency with which it can convert heat 
into useful work, 
and cannot be captured by a standard Born-Markov analysis of the system-reservoir interactions.  

\end{abstract}


\maketitle


In the same way that fundamental limitations such as the Carnot bound of classical 
engine cycles 
impose strict constraints on classical 
devices,
the fundamental limitations of quantum mechanical cycles 
set the stage for what is achievable with quantum devices. 
We now have 
a good understanding of quantum heat engines in the idealized situation of adiabatic dynamics and weak coupling to heat baths~\cite{0305-4470-12-5-007, :/content/aip/journal/jcp/80/4/10.1063/1.446862, 1367-2630-8-5-083, PhysRevE.65.055102,PhysRevE.76.031105, PhysRevLett.105.130401, PhysRevE.87.042131, PhysRevLett.112.030602, doi:10.1146/annurev, Skrzypczyk:2014aa, Gelbwaser-Klimovsky2015, doi:10.1021/acs.jpclett.5b01404, PhysRevB.91.195406, PhysRevX.5.031044, e18040124, PhysRevB.93.041418,PhysRevB.94.184503,PhysRevLett.119.170602,Marti2017,PhysRevA.99.052106, wiedmann19}.
Recent experimental realisations of nanoscale heat engines~\cite{Rosnagel2015,PhysRevLett.123.080602,PhysRevLett.122.110601,PhysRevLett.122.240602,PhysRevLett.123.240601}, however, operate well outside this regime,
and effects resulting from quantum coherence 
are known to impact a machine's efficiency. 
In particular, 
operating quantum thermodynamic cycles faster than adiabatically typically results in coherence generation during the work extraction strokes~\cite{PhysRevE.65.055102, Pekola181210933} with correspondingly reduced efficiencies. 
That this effect, termed quantum friction, 
can be avoided by dephasing (quantum lubrication~\cite{PhysRevE.73.025107}) or with control protocols such as shortcuts to adiabaticity~\cite{1212.6343,delcampo2015} is now well established. 
It thus implies a technical challenge, but not necessarily a fundamental limitation.

In contrast, the impact of quantum coherence entering the cycle via the \emph{heat exchange} strokes has not so far been considered. Primarily, this is because typical treatments of 
quantum 
engine cycles
assume idealized weak-coupling, 
in which system-reservoir correlations are not 
explicit. 
In the adiabatic regime where no coherence 
accumulates 
during the work extraction strokes, this 
assumption also results 
in a reduced system state that remains diagonal in its energy eigenbasis during the heat exchange strokes. Hence, the quantum and classical 
cycles become largely 
equivalent. 

Here we move beyond the limitations of the weak-coupling approach to consider the impact of 
system-reservoir correlations accrued as a result of finite coupling during the heat exchange strokes of a quantum Otto cycle. By remaining 
adiabatic along the work extraction strokes, we show that 
coherences resulting from system-reservoir correlations lead to performance losses even where conventional quantum friction plays no role. This constitutes a distinct finite-coupling quantum limit to non-equilibrium engine cycles that cannot be captured by a standard weak-coupling approach. Neither can it be avoided by straightforward generalisation of the techniques applied along the isentropic stokes. As we show, for finite system-reservoir coupling the appropriate basis for employing quantum lubrication comprises both system and reservoir components, and is thus difficult to identify experimentally. 
Likewise, control protocols are challenging to analyse even theoretically beyond the weak-coupling limit. 
Alternatively, dynamical decoupling could be employed to prevent the growth of system-reservoir correlations, but as this effectively decouples the system and reservoir, it would also simply stop the machine from working. Our work therefore reveals a crucial restriction on quantum engine performance, 
which is not yet clearly avoidable even in principle. 

We consider a two-level system (TLS) 
interacting separately with two heat reservoirs, at temperatures $T_h$ and $T_c$  ($T_h > T_c)$.
The 
Otto cycle 
consists of four strokes 
labelled by eight points: $A'BB'CC'DD'A$. 
\emph{Hot isochore}: at $A^{\prime}$ the TLS is coupled to the hot reservoir with which it interacts for a time $\tau_i$ to reach $B$. 
The interaction 
is then switched off instantaneously 
($B^{\prime}$).
\emph{Isentropic expansion}: the TLS Hamiltonian, $H_S$, is tuned over a time $\tau$ to reduce 
the gap between the two energy eigenvalues, 
reaching point $C$. 
Now, the 
TLS-cold reservoir interaction is switched on suddenly 
($C^{\prime}$).
\emph{Cold isochore}: the TLS interacts with the cold reservoir 
for a time $\tau_i$, to reach point $D$.
The system is then decoupled from the cold reservoir 
($D^{\prime}$). 
\emph{Isentropic compression}: $H_S$ is tuned for a time $\tau$ 
to increase the gap between energy eigenvalues 
back to 
the level 
at $A^\prime$, reaching point $A$. 
The cycle is completed by 
turning on the interaction with the hot reservoir 
($A^{\prime}$). 
These strokes are repeated until a limit cycle is reached \cite{Kosloff2017}. 

The full Hamiltonian 
reads 
\begin{equation} 
H(t)=H_{S}(t)+\sum_j(H_{R_j}
+H_{I_{j}}),
\label{Original_Ham} \end{equation}
where 
$H_{S}(t) = \epsilon(t)\sigma_{z}/2+\Delta(t)\sigma_{x}/2$, 
$H_{R_j} =  \sum_{k}\omega_{k_j}b_{k_j}^{\dagger}b_{k_j}$, 
and $H_{I_{j}} =  -\sigma_{z}\sum_{k_j}f_{k_j}(b_{k_j}^{\dagger}+b_{k_j})$ for $j=h,c$, denoting the hot ($h$) and cold ($c$) reservoir. 
Here, $\sigma_z$ and $\sigma_x$ 
are the usual TLS Pauli 
operators. 
The time dependence 
arises over the isentropic strokes when 
the splitting $\mu(t) = \sqrt{\epsilon(t)^2 + \Delta^2(t)}$ 
between the energy eigenlevels of $H_S(t)$ is tuned.
We label the values of $\mu$ during the hot or cold stage of the cycle 
as $\mu_h$ and $\mu_c$, respectively.
Bosonic reservoir annihilation operators for excitations at frequencies $\omega_{k_j}$ 
are given 
by $b_{k_j}$. 
The TLS-reservoir 
coupling is via $H_{I_j}$ 
with strengths $f_{k_j}$, 
and is present only during the relevant isochore.

Studies of quantum Otto cycles generally 
invoke the weak coupling
assumption, i.e.~that the interaction terms are (negligibly) small. This
leads to a tractable analysis in terms of a quantum state 
of the TLS and reservoirs approximated to remain 
in tensor product form 
at all times. In the
finite coupling regime of interest, where the interaction terms cannot be neglected,
it is a more involved task to compute the state 
around
the cycle~\cite{Marti2017,PhysRevA.99.052106, wiedmann19}. In order to access this regime, we extend 
the reaction-coordinate (RC) formalism
~\cite{AhsanReview, Iles-Smith2015, PhysRevA.90.032114,PhysRevE.95.032139, arXiv:1602.01340, arXiV:1711.00706} 
from the infinite time Otto cycle considered in Ref.~\cite{PhysRevE.95.032139} to inherently non-equilibrium, finite time cycles. In this approach, 
the 
interaction terms in Eq.~(\ref{Original_Ham}) are unitarily mapped to 
collective modes (the RCs) such that 
\begin{equation}\label{mappingHI} 
H_{I_j}=-\sigma_{z}\sum_{k_j}f_{k_j}(b_{k_j}^{\dagger}+b_{k_j})=-\lambda_j\sigma_{z}(a_j^{\dagger}+a_j), 
\end{equation}  
where 
$a_j$ annihilates an excitation in the RC mode for reservoir $j$ with natural frequency
$\Omega_j$, and 
$\lambda_j=({\sum_k f_{k_j}^{2}})^{1/2}$.
The reservoir Hamiltonians become
$H_{R_j}=
\Omega_j a_j^{\dagger}a_j +\sum_{k_j}g_{k_j}(a_j^{\dagger}+a_j)(r_{k_j}^{\dagger}+r_{k_j})
+\sum_{k_j}\nu_{k_j}r_{k_j}^{{\dagger}}r_{k_j}$, 
while the system Hamiltonian remains unchanged. Here, 
$r_{k_j}$ annihilates 
an excitation at frequency $\nu_{k_j}$ in 
a redefined residual reservoir
which interacts weakly with the corresponding RC mode via couplings $g_{k_j}$.
The full Hamiltonian becomes 
$H (t) = H_{S^\prime}(t) + \sum_jH_{R^\prime_j}$, 
where 
$H_{S^\prime}(t)  =  {\epsilon(t)}\sigma_{z}/2+{\Delta(t)}\sigma_{x}/2-\sum_j\lambda_j\sigma_{z}(a_j^{\dagger}+a_j)
+\Omega_j a_j^{\dagger}a_j$, and 
$H_{R^\prime_j}  =  \sum_{k_j}g_{k_j}(a_j^{\dagger}+a_j)(r_{k_j}^{\dagger}+r_{k_j})+\sum_{k_j}\nu_{k_j}r_{k_j}^{\dagger}r_{k_j}$. 
The RC mapping involves an enlarged view of a redefined 
system $S^\prime$
whose self-energy $H_{S^\prime} (t)$ now additionally incorporates 
the self-energies 
of the reservoir RCs as well as the TLS-RC coupling terms. 
The remaining terms 
represent residual environments $R^\prime_j$ 
and their interactions with the corresponding RC, which may be treated as Markovian~\cite{Iles-Smith2015, PhysRevA.90.032114}. 
Nevertheless, through the RCs, correlations between the TLS and the 
reservoirs 
will form and indeed persist even in the limit cycle. 

In the original representation 
we characterise the 
TLS-reservoir interactions 
via 
spectral densities  
$J_j(\omega)\equiv\sum_k {f^2_{k_j}} \delta(\omega - \omega_{k_j})={\alpha \omega \omega_{c}}/{({\omega}^{2}+{\omega}_{c}^{2})}$, taken to be the same for each reservoir,  
with coupling strength $\alpha$ and cutoff frequency $\omega_{c}$. 
As the residual reservoirs
are traced out when deriving a master equation for the enlarged system
$S'$, to determine the RC mapping one simply needs to find 
the spectral density $\tilde{J_j}(\nu) \equiv \sum_k {g^2_{k_j}} \delta(\nu - \nu_{k_j})$ 
that characterizes the coupling between $S'$ and the residual 
reservoirs, as well as 
$\Omega_j$ and $\lambda_j$,
such that the Heisenberg equations of motion for operators in the TLS subspace are equivalent
in both pictures. Imposing this results
in 
$\Omega_j= 2 \pi \gamma \omega_c$, 
$\lambda_j = \sqrt{{\pi \alpha \Omega_j}/{2}}$, 
and 
$\tilde{J}_j(\nu)= \nu{\sqrt{\epsilon^2 + \Delta^2}}/{2 \pi \omega_c}$~\cite{PhysRevA.90.032114}. 
The master equation governing the dynamics of $S^\prime$ during the $j$ ($=h,c$) isochore is 
$\dot{\rho}(t) = L_j[\rho(t)]$, 
with
$L_j[\rho(t)] \equiv - i [ H_{S'_j} , \rho(t)] - [ A_j,[ \chi_j, \rho(t) ] ] 
 + [ A_j,\{ \Xi_j, \rho(t) \} ]$ and $\rho(t)$ the state of the TLS plus both RCs. 
Here, 
$A_j = a_j + a_j^{\dagger}$, the self-Hamiltonian $H_{S^\prime_j}$ only includes interaction terms for the $j$ RC, 
$\chi_j = \gamma \int^\infty_0 d\tau \int^\infty_0 d\omega \omega \cos (\omega \tau) \coth \left(\frac{\beta_h \omega}{2}\right) A_j(-\tau)$, and 
$\Xi_j = \gamma \int^\infty_0 d\tau \int^\infty_0 d\omega \cos (\omega \tau) \left[H_{S_j^\prime}, A_j(-\tau) \right]$, 
for $A_j(\tau) \equiv e^ {i H_{S'_j} \tau}A_j e^{-i H_{S^\prime_j} \tau}$~\cite{PhysRevA.90.032114}. 

Thermodynamic treatments of the Otto cycle usually consider thermal reservoir resources. 
We 
wish 
to isolate strong-coupling effects from any 
due to coupling to heat reservoirs that are out of 
equilbrium at the start of each isochore. To this end, we include a mechanism in our finite time cycle 
to ensure that any uncoupled reservoir returns to thermal equilibrium by the time it is coupled to the TLS once more. 
For the 
uncoupled reservoir $j$, which has been driven out of equilibrium during the previous 
isochore, to return to thermal equilibrium at temperature $T_j$ while the 
other reservoir and TLS interact, 
we can add terms to the master equation 
that act only on the uncoupled 
RC and hence do not depend on the full system plus RC eigenstructure. The uncoupled RC is a harmonic oscillator so we 
add standard 
dissipators 
$L_{d_j} [\rho(t)] =  \gamma_{d} (N_j+1){\mathcal L}_{a_j} [\rho(t)] +  \gamma_{d} N_j{\mathcal L}_{a^{\dagger}_j} [\rho(t)]$,
where ${\mathcal L}_{O}[\rho(t)] = O \rho(t) O^\dagger - \frac{1}{2}\{O^\dagger O,\rho(t)\}$ and $N_j = (e^{\beta_j \Omega_j}-1)^{-1}$, 
with $\beta_j=1/T_j$~\cite{Breuer:2002aa, Carmichael:1999aa} . 
We choose 
$\gamma_{d}$ to ensure thermalisation occurs over a timescale such that the TLS re-couples to a thermal reservoir at the start of the subsequent isochore. 
The cycle work output 
is given by the net energy change of the system across each of the isentropic strokes. For strong coupling, this also involves accounting for the energetic costs associated with turning off the interaction term at the end of the isochores; 
there are no coupling costs due to rethermalisation of the RCs when uncoupled. 
This leads to 
work output~\cite{PhysRevE.95.032139}
\begin{eqnarray} 
W = & & \textrm{tr}\left[H_{S}^{c}\rho^{C}\right] - \textrm{tr}\left[H_{S}^{h}\rho^{B}\right]  + \textrm{tr}\left[H_{S}^{h}\rho^{A}\right] \nonumber \\
&-& \textrm{tr}\left[H_{S}^{c}\rho^{D}\right] - \textrm{tr}\left[H_{I_{h}}\rho^{B}\right]-\textrm{tr}\left[H_{I_{c}}\rho^{D}\right], 
\label{Workfinite}
\end{eqnarray}
where $h$ and $c$ superscripts 
indicate that the TLS splitting 
is set to $\mu_{h}$ and $\mu_{c}$, respectively. The density operator $\rho$ is labelled with superscripts $A - D$ indicating the various points around the cycle and as before represents the state of the enlarged $S^\prime$. 
The energy transferred into the system during the hot isochore is given by
$Q = \textrm{tr}[H_{S'_h}^{h}\rho^{B}] - \textrm{tr}[H_{S'_h}^{h}\rho^{A}]$, 
and we shall term this heat. 
Through $H_{S'_h}$ this expression contains contributions from the TLS-hot RC interaction energy 
as well as 
from the hot RC being pulled out of equilibrium, both of which 
are neglected in weak-coupling treatments. 

To evaluate the heat and net work output
we need to compute the 
states $\rho^{A-D}$ of the enlarged system $S^\prime$ 
when the engine is operating in the limit cycle. In the infinite time (equilibrium) version of the cycle, this involves taking the steady state solution of 
the RC master equation 
along each isochore, 
followed  by unitary evolution along the isentropic strokes. For the non-equilibrium case 
considered here, the calculation is more involved and we must numerically solve the dynamical equation of motion for the full state of $S^\prime$ for a particular isochore time $\tau_i$. 

\begin{figure}[t]
\centering
  \includegraphics[scale=0.220]{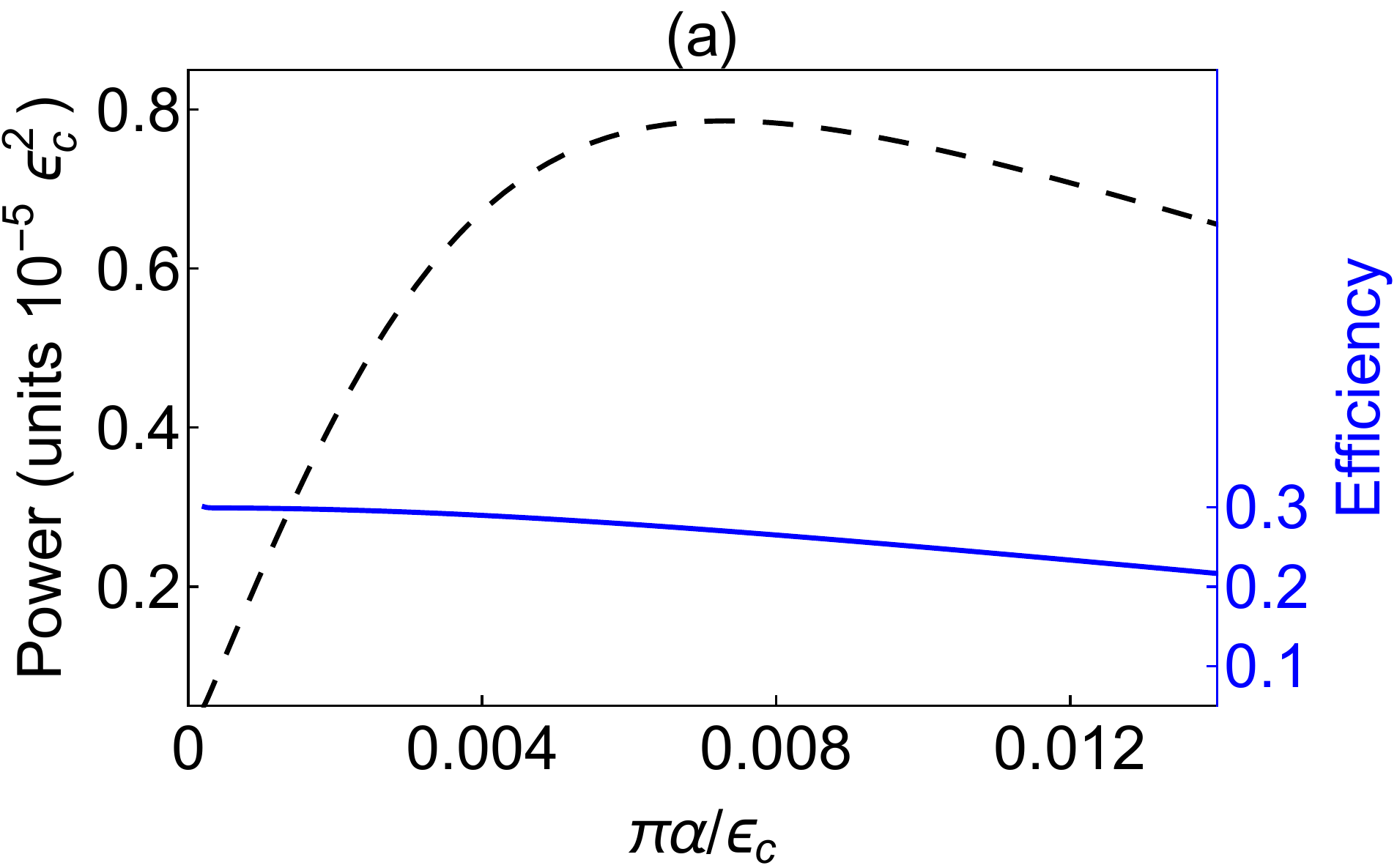}
    \includegraphics[scale=0.220]{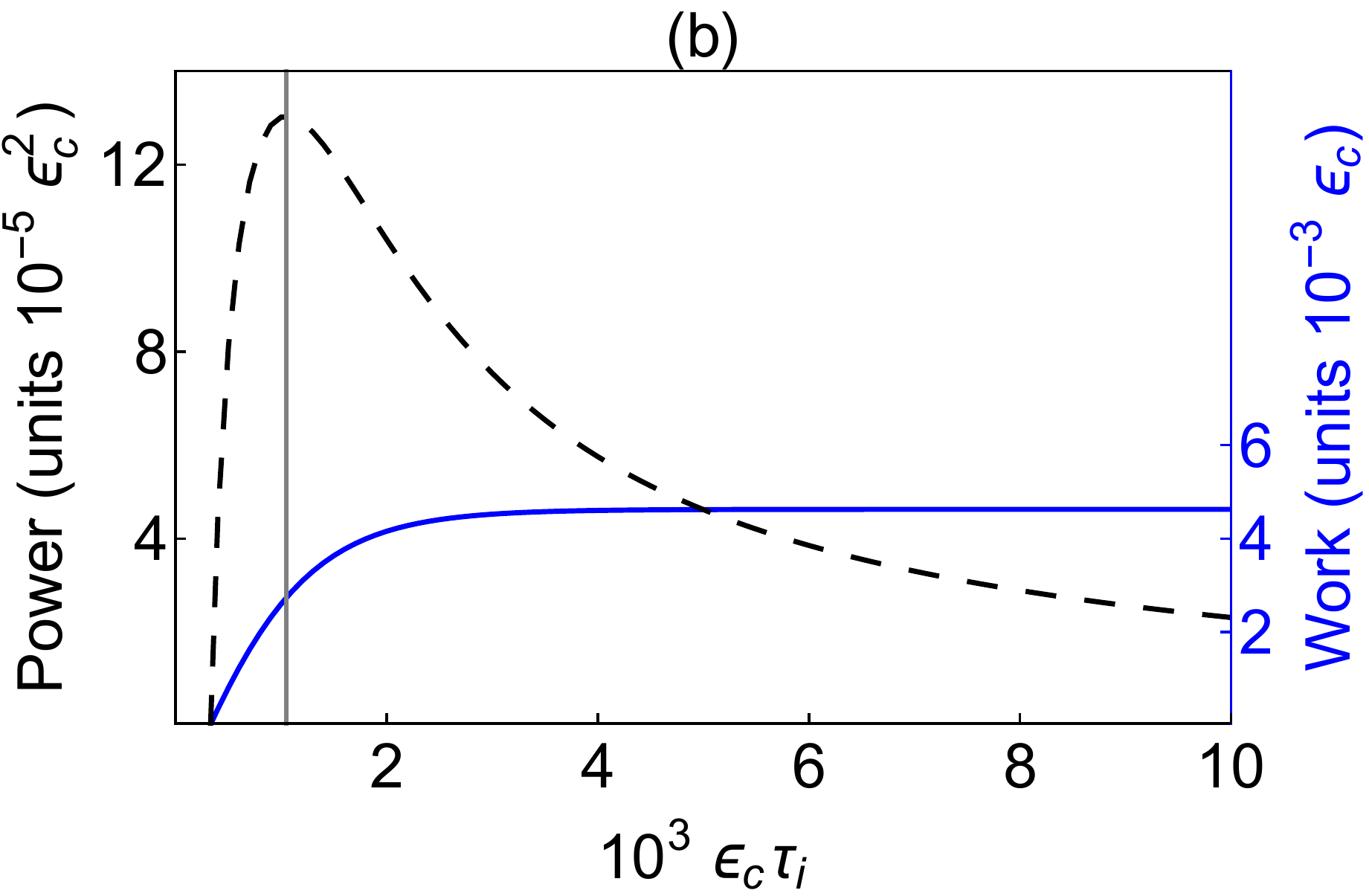}
\caption[Non-equilibrium heat engine metrics as a function of coupling strength]{
(a) Power output (black dashed) and efficiency (blue solid) of the non-equilibrium TLS quantum Otto cycle, each as a function of coupling strength $\pi\alpha/\epsilon_c$ for $\epsilon_c\tau_i = 3000$. (b) 
Power (black dashed) and work (blue solid) output against isochore time for $\alpha = 0.01\epsilon_c / \pi$. 
Parameters: 
$\epsilon_h = 1.5\epsilon_c$, $\Delta_c = \epsilon_c$, $\Delta_h = 1.5\epsilon_c$, $\omega_c = 0.265\epsilon_c$, $\epsilon_c\beta_h = 0.95$, $\epsilon_c\beta_c = 2.5$, and $9$ states in each RC.}
\label{metrics_coupling}
\end{figure}

During the isentropic strokes we tune $\epsilon(t)$ and $\Delta(t)$ such that $(\epsilon_h,\Delta_h) \leftrightarrow (\epsilon_c,\Delta_c)$, 
with [$H_S(t), H_S(t')]=0$. This ensures that the TLS Hamiltonians at the start of the stroke, $H_S(0)$, and at the end, 
$H_S(\tau)$, share a common energy eigenbasis. The TLS quantum state 
then adiabatically follows the change in splitting and no coherence develops 
even when the stroke is carried out in a finite time $\tau$. We thus avoid any quantum friction along the isentropic strokes, and any variations in performance due to the generation of coherence can instead be attributed to the isochores. For the purposes of maximising power output, it is then preferable to complete this stroke quickly and we consider the limit $\tau \rightarrow 0$. 

In Fig.~\ref{metrics_coupling} (a) we plot the engine's power output and efficiency 
at finite coupling. 
The power output initially increases with coupling strength until a turnover is reached 
as reservoir decoupling costs begin to dominate over the increase in work output. Note that we are considering here a finite isochore time, $\tau_i$, shorter than that necessary for a stationary state to be reached along the isochores. 
The energy absorbed from the hot reservoir increases with coupling strength too and this leads to an engine efficiency which decreases monotonically with coupling strength. 
A similar finding is also reflected in the analysis of the strong coupling regime for a heat engine in Ref.~\cite{Marti2017} but there the isochores are long enough that the engine is considered to have reached arbitrarily close to equilibrium. In the present case, we show that in fact power output can be maximised before this equilibrium has been reached in Fig.~\ref{metrics_coupling} (b). 
Here we plot the 
work and power outputs as a function of the isochore time $\tau_i$ for an intermediate coupling strength. 
As the system $S^{\prime}$ approaches equilibrium, the work output saturates. Power output, however, is maximised before this equilibrium is reached. 
If the desired engine metric 
is how much power can be produced, it is thus preferable to operate out of equilibrium, by choosing shorter isochore times and intermediate coupling strengths.

\begin{figure}
\centering
  \includegraphics[scale=0.220]{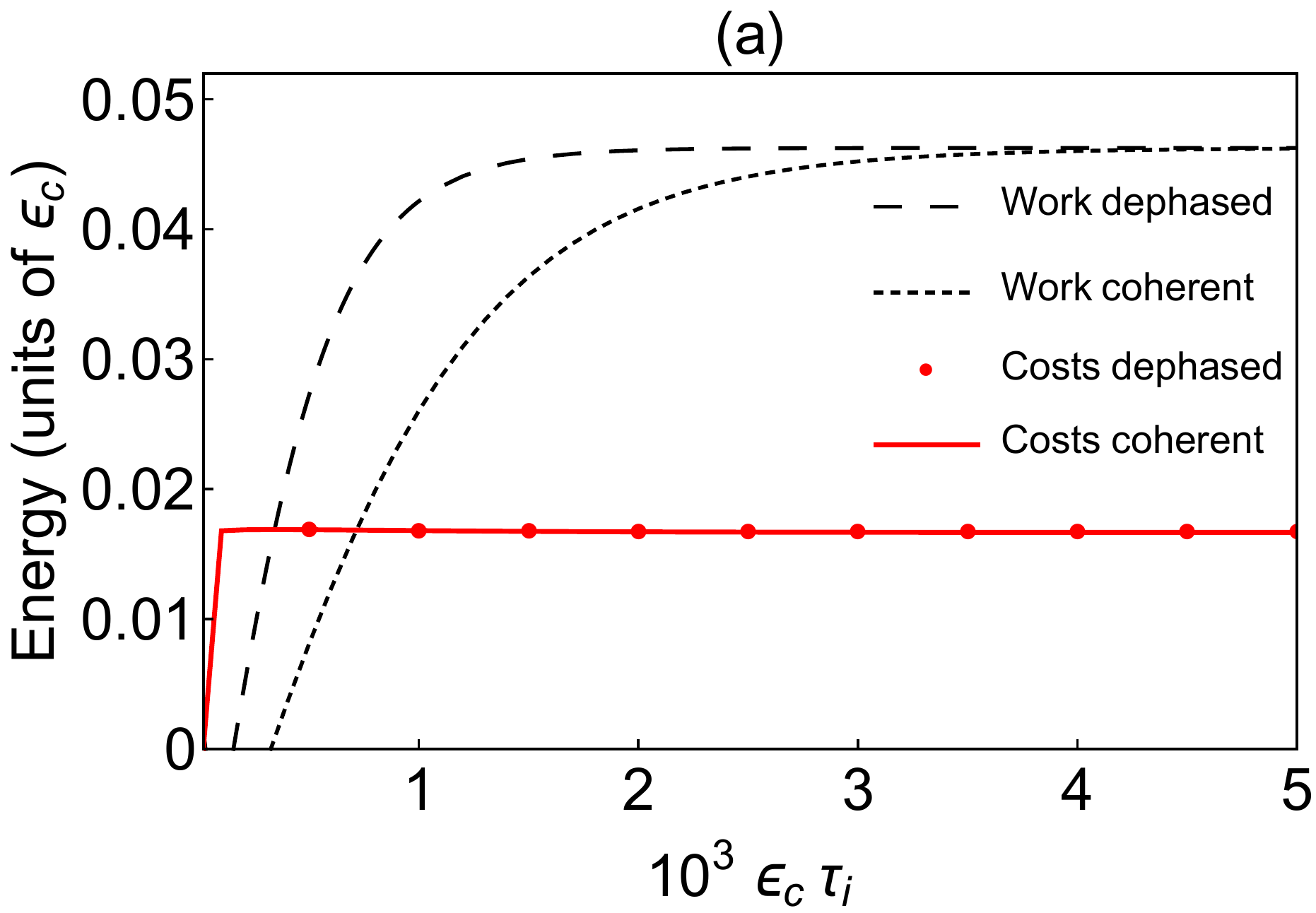}
    \includegraphics[scale=0.220]{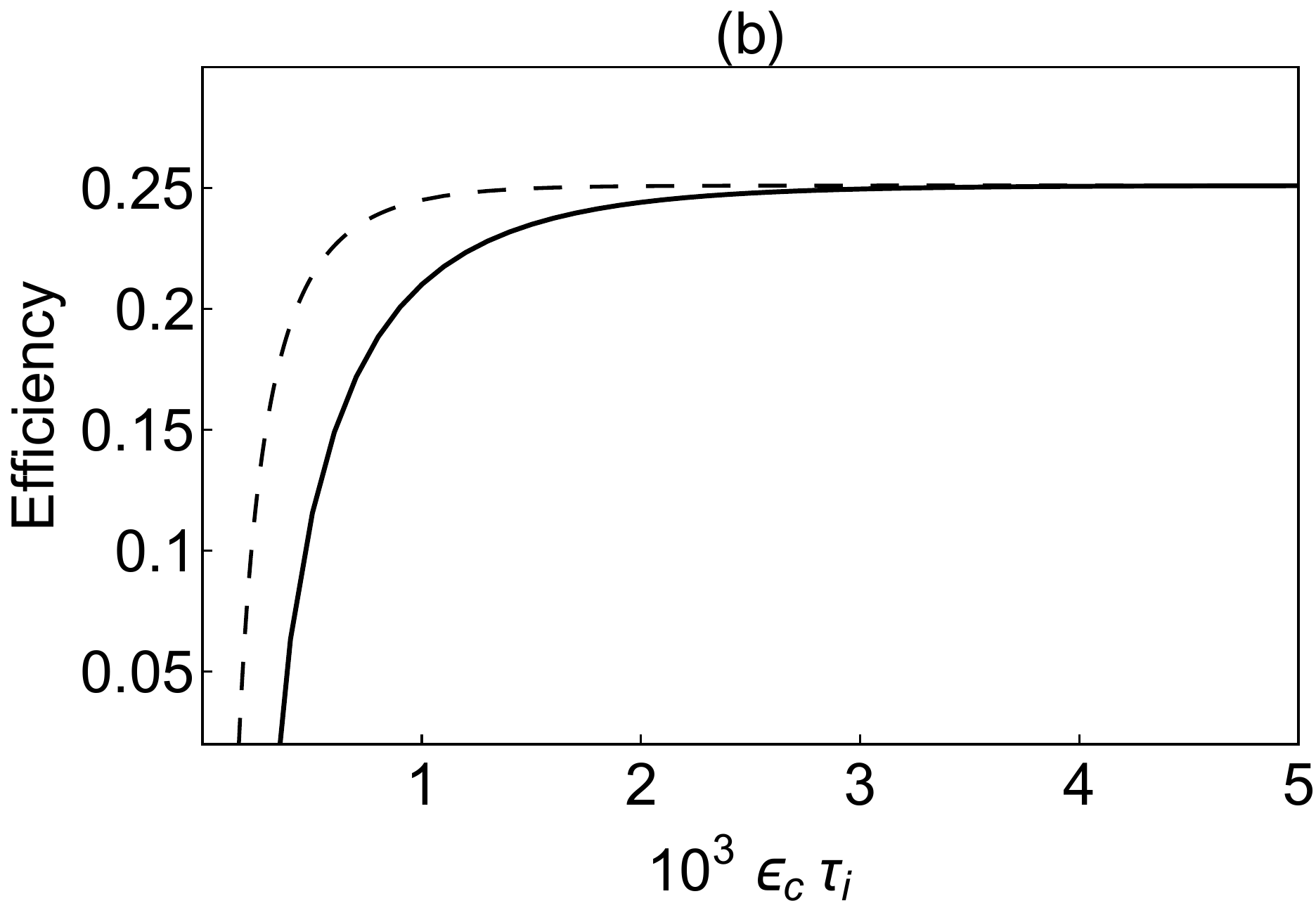}
\caption[Coherent and incoherent engine cycles]{
Work and decoupling costs (a) and efficiency (b) plotted against isochore time $\epsilon_c\tau_i$. 
In (b) the solid curve represents the fully coherent engine, the dashed curve depicts the incoherent engine.  
Parameters: $\alpha=0.01\epsilon_c/\pi$, with others as in Fig.~\ref{metrics_coupling}.}
\label{Coherent_dephased_figures1}
\end{figure}

We now wish to explore the quantum nature of the heat engine, and specifically isolate the effects on engine performance of system-reservoir correlations accrued during the (heat exchange) isochoric strokes as a result of finite coupling with the reservoirs. These correlations manifest themselves in finite coherences in the working system state. We shall therefore make a distinction between a fully quantum version of the cycle, where system-reservoir correlations lead to quantum coherence being generated along the isochores, and one where coherence is prevented from accumulating. In this latter version of the cycle, we introduce into the master equation 
terms that induce pure dephasing \cite{Breuer:2002aa, Schlosshauer:2007aa, TEMPEL2011130, 1367-2630-11-3-033003,PhysRevE.73.025107} in the energy eigenbasis of the working system, while taking care that these have no energetic contribution to the overall evaluation of work output or energy exchange with the reservoirs.
To meet with these criteria, the pure dephasing terms must commute with the mapped Hamiltonian $H_{S^{\prime}}$, which we diagonalise and write as 
$H_{S^\prime} = \sum_n E_n \ket{E_n}\bra{E_n}$.
We then construct a pure dephasing Liouvillian 
\begin{equation}
L_{dep}[\rho(t)] = \gamma_{dep} \sum_n \big[\ket{E_n}\bra{E_n},[\ket{E_n}\bra{E_n}, \rho(t) ]\big].
\end{equation}
We are free to choose the value of $\gamma_{dep}$ to ensure dephasing occurs on an appropriate timescale, such that coherence is prevented from developing during the isochoric strokes of the engine. We can then compute the states of the working system at various points around the cycle as previously described, but with the addition of these terms to 
both isochores. 
We define this as the incoherent engine. 
We stress that these terms achieve pure dephasing in the energy eigenbasis of the enlarged TLS plus RCs (i.e.~$H_{S'}$) rather than just the TLS energy eigenbasis. This is the natural choice for a system interacting strongly with an environment. Introducing pure dephasing terms 
that only act on the TLS would be inappropriate since they do not commute with the unitary part of the master equation, 
and 
thus have a non-zero energetic contribution. 


\begin{figure}
\centering
  \includegraphics[scale=0.4]{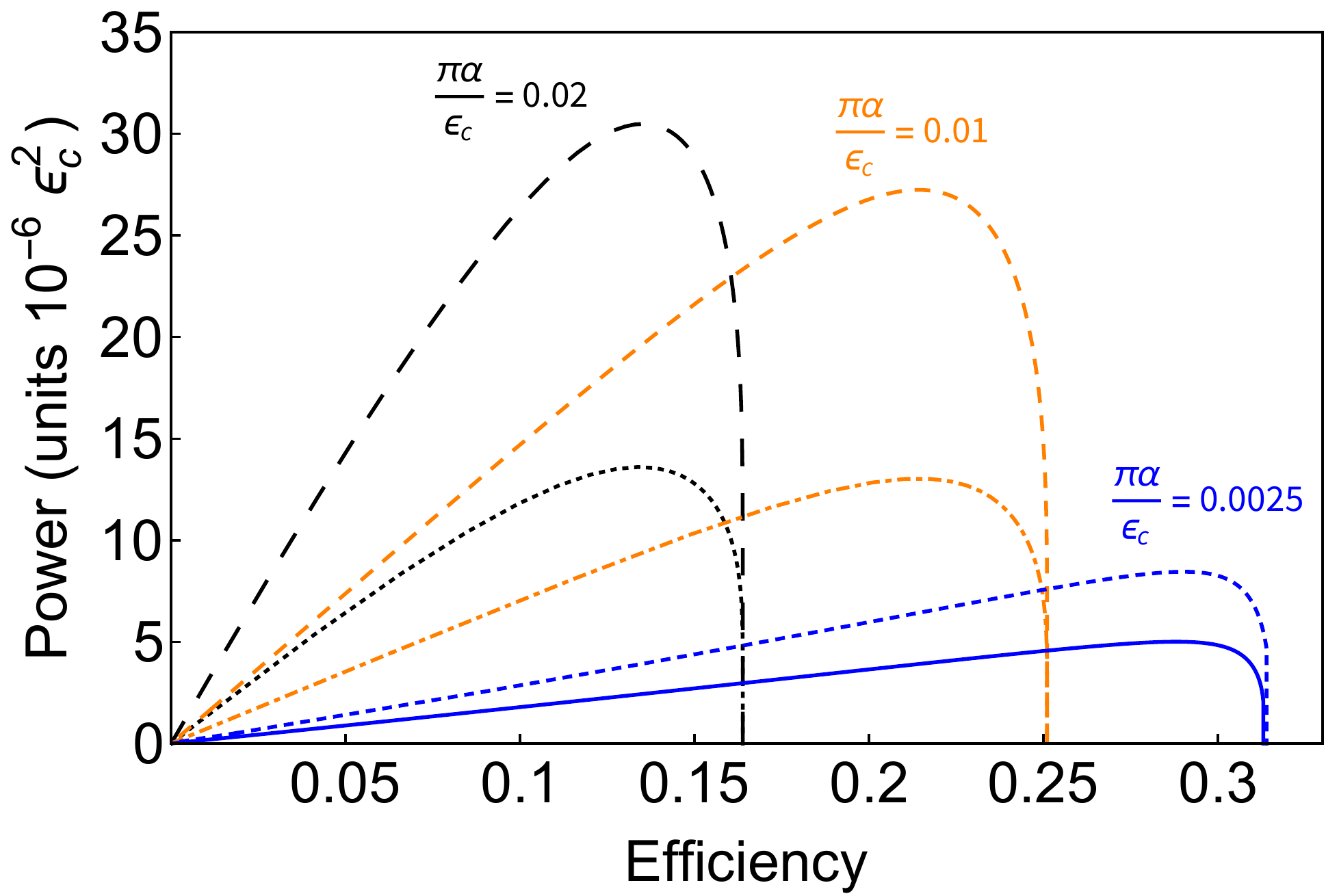}
\caption[Power versus efficiency optimisation in coherent and incoherent engines]{Parametric plot of power output against efficiency for 
various 
coupling strengths $\alpha$. The dashed curves represent the incoherent engine. The dotted, dot-dashed, and solid curves depict the fully coherent engine. The parameter varied along the curves is the isochore time $\tau_i$. Other parameters as in Fig.~\ref{metrics_coupling}.}
\label{Coherent_dephased_figures2}
\end{figure}

We compare these two versions of the Otto cycle 
and analyse the effect of quantum coherence on engine performance in 
Fig.~\ref{Coherent_dephased_figures1} as a function of isochore time $\tau_i$. 
At large $\tau_i$, the incoherent and coherent engines converge. Here, the state of $S^\prime$ approaches thermal equilbrium with the relevant residual reservoir which is maintained at the hot or cold temperature. This state is then diagonal in the energy eigenbasis of $S^\prime$, and so no coherence is present in either type of engine at points $B$ or $D$ if the isochore is long enough. For shorter times, however, this is not the case, and pure dephasing does have an appreciable effect on the engine metrics. The dephased engine absorbs more heat along the hot isochore (not shown) but outputs a net work large enough to compensate, yielding a higher efficiency at short times than the fully coherent engine. Decoupling costs remain comparable. These are dominated by the RC being driven out of equilibrium along the isochores, which is not prevented by dephasing. In Fig.~\ref{Coherent_dephased_figures2} we show this improved engine performance in parametric plots of power versus efficiency 
for a selection of coupling strengths. Even at weaker but finite coupling, where coherence generation is less severe, an improvement 
over the coherent engine can be achieved by dephasing. 

\begin{figure}
\centering
  \includegraphics[scale=0.2]{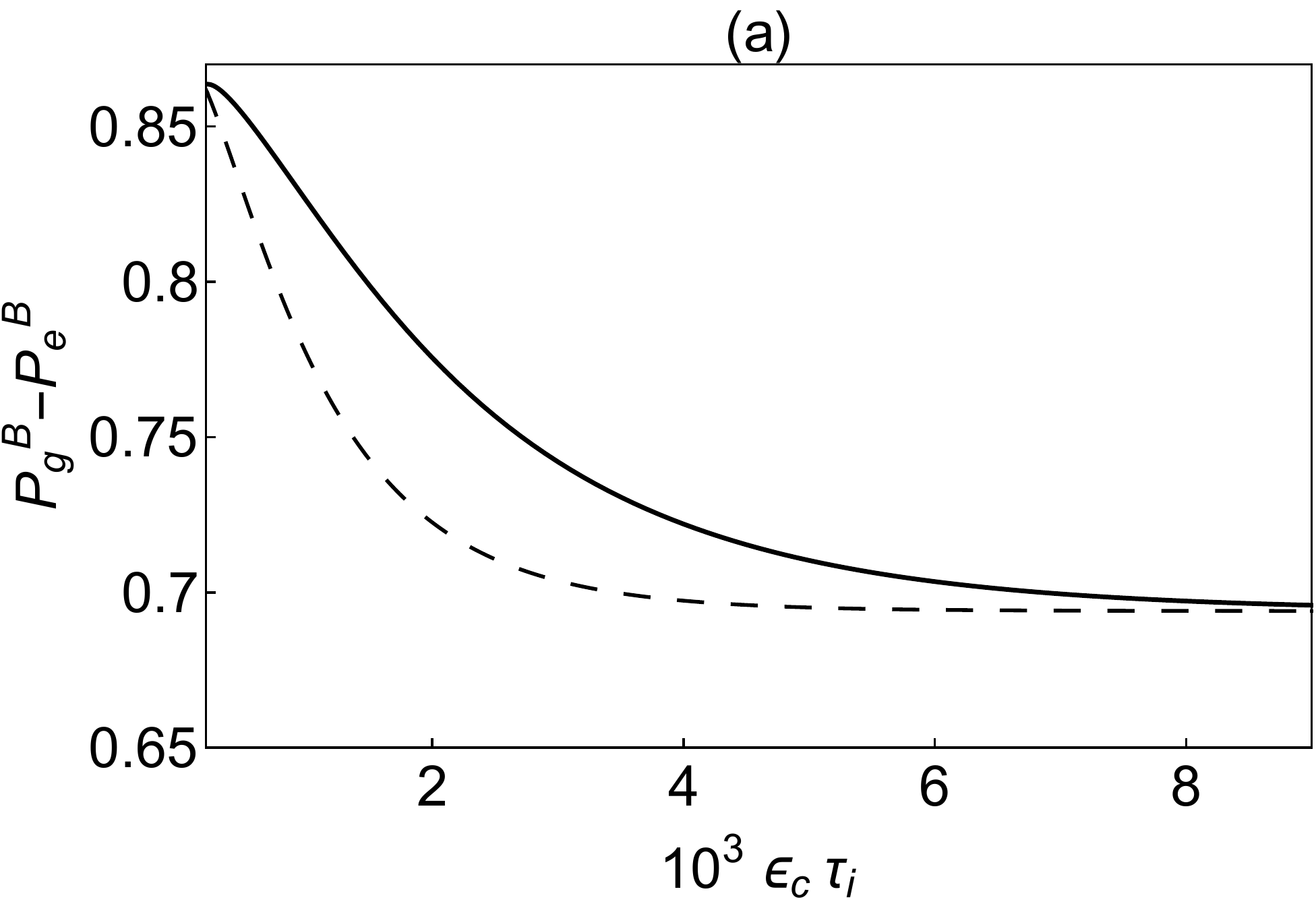}
 \includegraphics[scale=0.2]{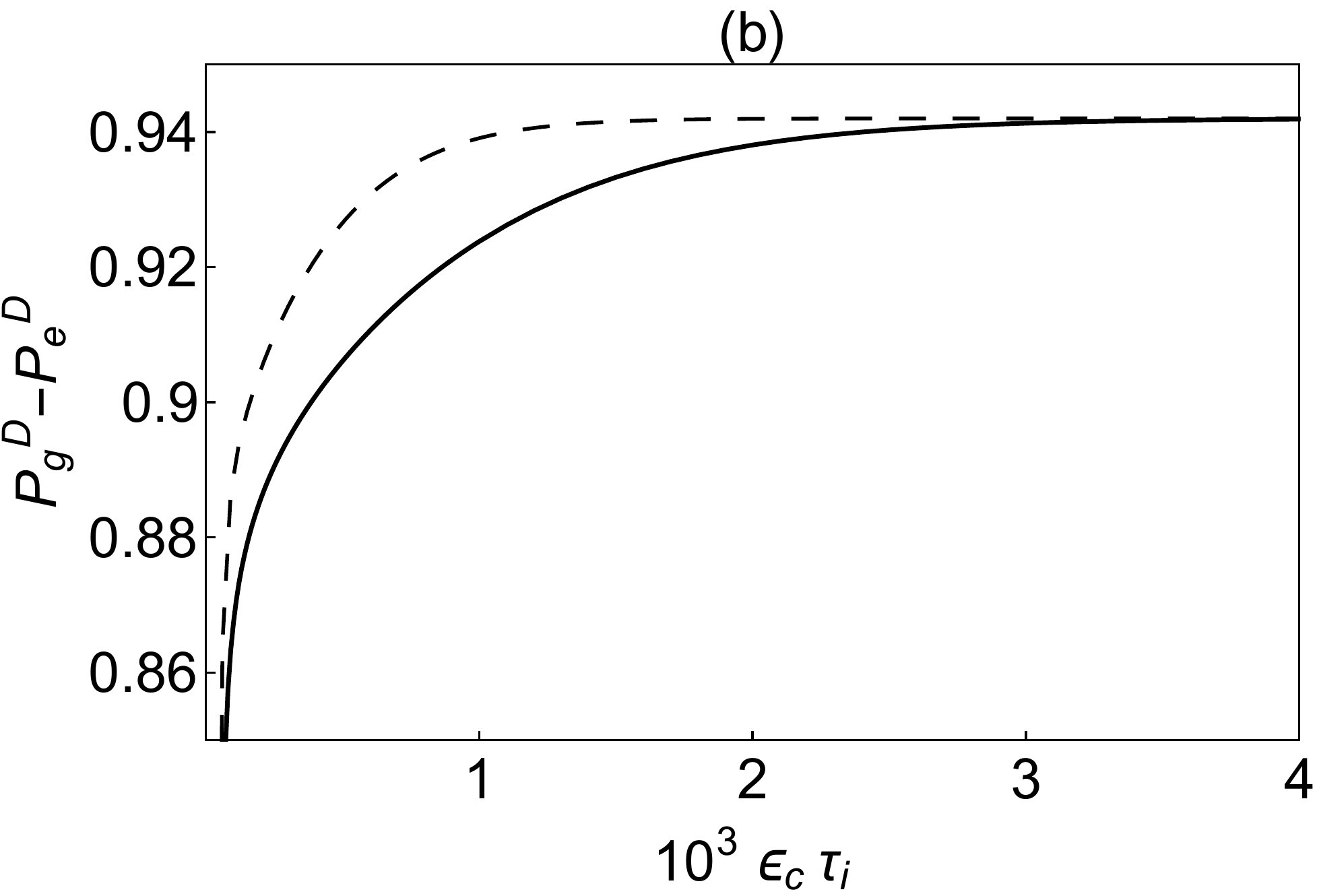}
  \includegraphics[scale=0.2]{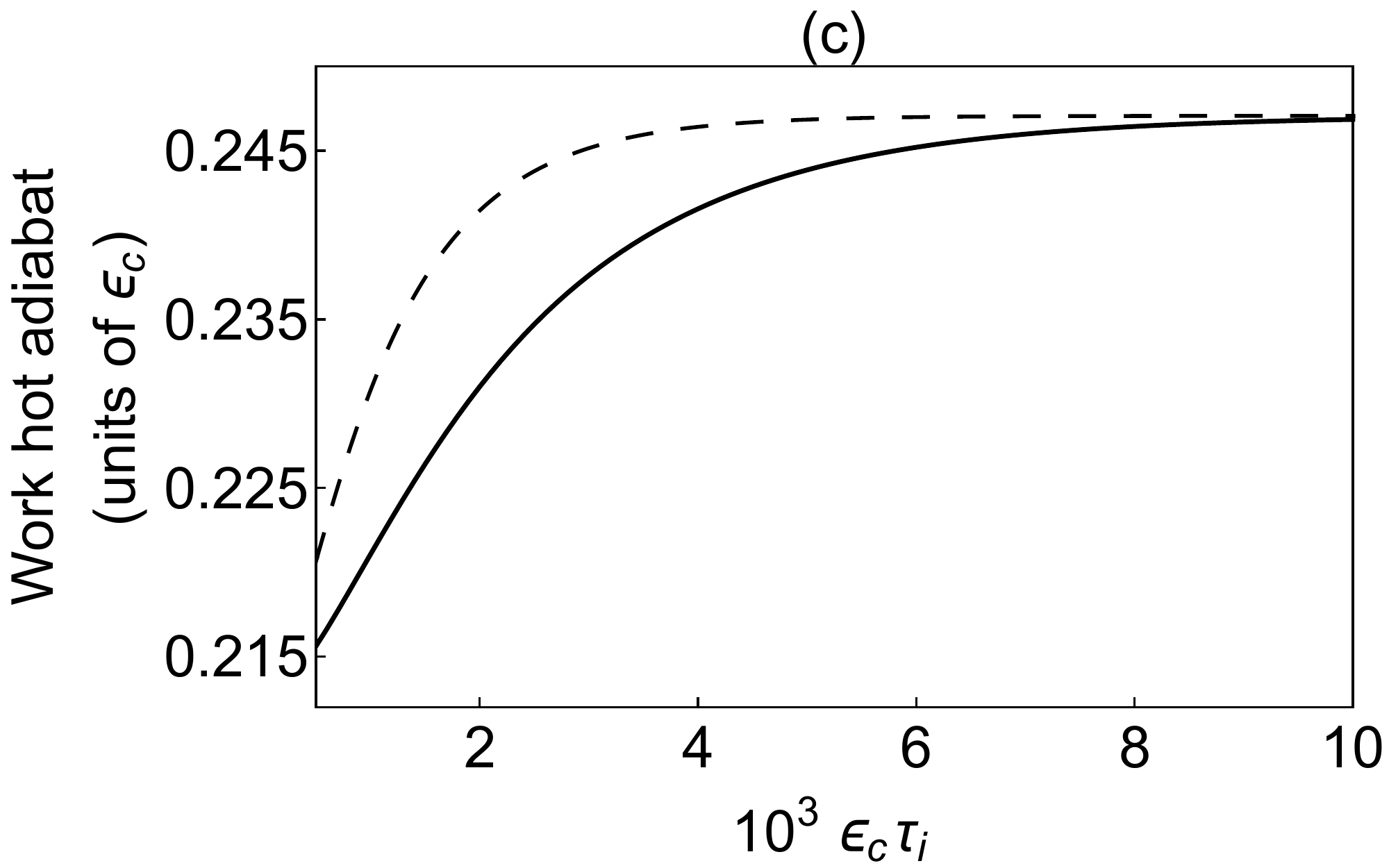}
  \includegraphics[scale=0.2]{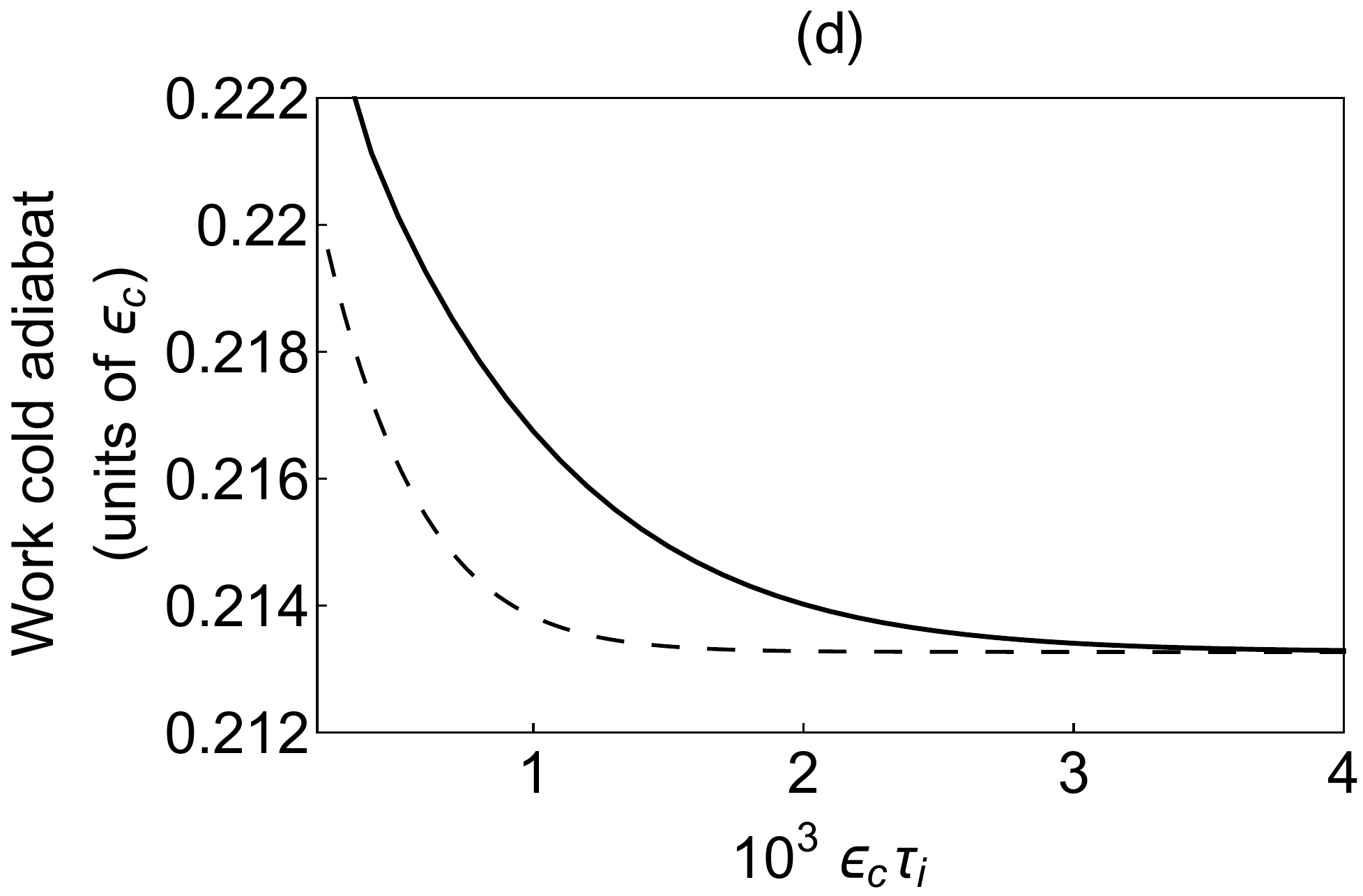}
\caption[The effect of dephasing on engine populations and adiabatic strokes]{
TLS eigenstate population difference at point B in the cycle (a), 
and at point D (b); 
work extracted during the hot adiabatic stroke (c) and work input during the cold adiabatic stroke (d), plotted against isochore time $\tau_i$. Solid curves represent the fully coherent engine, dashed curves depict the 
incoherent engine. Parameters as in Fig.~\ref{Coherent_dephased_figures1}.}
\label{Coherent_dephased_figures3}
\end{figure}

In a 
weak coupling treatment of the Otto cycle 
the TLS equilibrates to a Gibbs state for long isochore times, which is diagonal in the energy eigenbasis. 
With the isentropic strokes carried out adiabatically, populations in these energy eigenstates are kept constant and no coherence accumulates in the working system state. In fact, if the isentropic strokes remain adiabatic when the cycle is treated at finite time (as we consider), then no coherence is present during the weak-coupling cycle at any time. 
At non-negligible system-reservoir interaction strengths, however, correlations that are generated between the TLS and reservoirs  
lead to quantum coherence in the energy eigenbasis at the end of the isochores. This coherence will, in general, persist during the isentropic strokes even if they are performed adiabatically. In this way, the strongly coupled TLS Otto cycle is inherently quantum in nature.

We further illustrate how dephasing in the TLS plus RC basis accelerates the process of equilibration 
in Fig.~\ref{Coherent_dephased_figures3}. Of crucial importance for 
work calculations 
are the TLS population differences 
in the eigenbasis of $H_S$ [see Eq.~(\ref{Workfinite})]. 
In Fig.~\ref{Coherent_dephased_figures3} we show how these are affected by dephasing 
at point B, just before the hot adiabatic stroke begins, and at point D, just before the cold adiabatic stroke begins. 
In a weak-coupling analysis 
the TLS  populations would be completely insensitive to pure dephasing (which for weak coupling would be defined relative to $H_S$ to posses no energetic contribution). In stark contrast, for the finite coupling theory presented here, TLS populations are significantly affected by pure dephasing in the correct strong-coupling basis of $H_{S'}$. 
Specifically, the dephased engine displays a faster approach to the steady state 
than the fully coherent engine. At point B there is therefore a smaller population difference (a) at shorter $\tau_i$ and at point D a larger population difference (b). 
This results in greater work extracted during the hot adiabat (c) and smaller work invested during the cold adiabat (d) for the dephased engine. The reason is that 
at point B we wish to have as much population as possible in the excited state  
as this implies a 
higher 
(effective) TLS temperature. 
This entails that more heat has been absorbed from the hot reservoir during the hot isochore and that the subsequent hot adiabatic stroke can extract a larger amount of work. 
At point D, on the other hand, 
it is desirable to have as little population in the excited state as possible.  
This corresponds to a lower temperature, 
which means that as much heat as possible has been dissipated into the cold reservoir and that on the subsequent adiabatic stroke, the work performed on the system can be kept to a minimum. 



To summarise, 
we have shown that finite system-reservoir coupling imposes a 
bound on quantum heat engine performance that is not captured by standard weak-coupling 
approaches. 
Our results highlight that even when work extraction is adiabatic, 
quantum coherence enters the cycle through finite system-reservoir interactions. 
Protocols designed to avoid quantum friction along the work strokes are not straightforwardly extendable to suppress the generation of coherence along the isochores. We thus expect our findings to be of broad relevance to practical realisations of quantum engine cycles, 
and to stimulate 
a concerted effort to devise schemes to mitigate this quantum disadvantage.
\begin{acknowledgments}
We thank Luis Correa and Ronnie Kosloff for discussions. D.N. is supported by the EPSRC and the CDT in Controlled Quantum Dynamics. A.N. is supported by the EPSRC, Grant No. EP/N008154/1.
\end{acknowledgments}

%

\end{document}